\documentclass[a4paper,11pt]{article}
\pdfoutput=1 

\usepackage{jinstpub} 

\usepackage{xcolor}
\usepackage{siunitx}
\usepackage[inline]{enumitem}
\usepackage{textgreek}
\usepackage{xfrac}
\usepackage{systeme}

\graphicspath{{./images/}}
 
\title{Validating fast-ion wall-load IR analysis-methods against W7-X NBI empty-torus experiment}

\author[a,1]{S. \"{A}k\"{a}slompolo,\note{Corresponding author.}}
\author[a]{P. Drewelow}
\author[b]{Y. Gao}
\author[a,2]{A. Ali\note{Current affiliation: Max Planck Institute of Molecular Physiology, Germany}}
\author[c,3]{O. Asunta\note{Current affiliation: Tokamak Energy Ltd, UK}}
\author[a]{S. Bozhenkov}
\author[a]{J. Fellinger}
\author[a]{O.P. Ford}
\author[d]{N. den Harder}
\author[a]{D. Hartmann}
\author[a]{M. Jakubowski}
\author[a]{P. McNeely}
\author[a]{H. Niemann}
\author[e]{F. Pisano}
\author[a]{N. Rust}
\author[a]{A. Puig Sitjes}
\author[f]{M. Sleczka}
\author[a]{A. Spanier}
\author[a]{R.C. Wolf}

\affiliation[a]{Max-Planck-Institut f\"{u}r Plasmaphysik, Teilinstitut Greifswald,\\Wendelsteinstra\ss{}e 1, 17491 Greifswald Germany}
\affiliation[b]{Forschungzentrum J\"{u}lich, IEK-4\\52425 J\"{u}lich, Germany}
\affiliation[c]{Aalto University,\\00076 Aalto, Finland}
\affiliation[d]{Max-Planck-Institut f\"{u}r Plasmaphysik,\\Boltzmannstra\ss{}e 2, 85748 Garching, Germany}
\affiliation[e]{University of Cagliari,\\Via Universit\`{a}, 40 - 09124 Cagliari, Italy}
\affiliation[f]{University of Szczecin,\\al. Papie\.{z}a Jana Paw\l{}a II 22a, 70-453 Szczecin, Poland}

\emailAdd{simppa.akaslompolo@alumni.aalto.fi}

\abstract{
The first neutral beam injection (NBI) experiments in Wendelstein 7-X (W7-X) stellarator were conducted in the summer of 2018. 
The NBI system is used to heat the magnetically confined plasma by neutralising an accelerated hydrogen ion beam and directing it into the plasma, where the resulting energetic ions release their energy to heat the plasma.
The modelling of the NBI fast ion experiments has commenced, including estimation of the shine-through and the orbit-losses. 
The stellarator has a wide-angle infra red (IR) imaging system to monitor the machine plasma facing component surface temperatures.
This work validates the NBI model \emph{"Beamlet Based NBI (BBNBI)"} and the newly written synthetic IR camera model. The validation is accomplished by comparing the measured and the synthetic IR camera measurements of an experiment where the NBI was injected into the vacuum vessel without a plasma.
A good qualitative and quantitative match was found.
This agreement is further supported by spectroscopic and calibration measurements of the NBI and and IR camera systems.

}


\collaboration[c]{and the W7-X Team}

\begin{document}
\maketitle
\flushbottom

\section{Introduction}
\label{sec:intro}

The Wendelstein 7-X (W7-X) stellarator started operations in 2015 and the first neutral beam injection (NBI) experiments took place in 2018. 
W7-X is an optimised stellarator, with expected good fast ion confinement at high plasma pressure.
Demonstrating this is one of the high-level goals of the W7-X project, and is only possible with validated fast ion modelling tools.
In 2018, only few dedicated fast ion diagnostics were available.
However, the general purpose  wide-angle divertor infrared (IR) monitoring system~\cite{IRimagingSys4wallProtectionInW7X} is an effective tool to study NBI fast-ion orbit-losses~\cite{NBIwallLoadsInW7X} on the walls.
Modelling of the experiments with ASCOT4~\cite{ascot4ref} is in progress, but the modelling requires a validation of the synthetic IR model as well as a beamlet based NBI model (BBNBI)~\cite{Asunta14_BBNBI_reference}, and this article reports the validation work.
The physics of plasma facing component (PFC) heating up due to orbit loss fast ions is the same as heating due to the fast neutrals. Hence, the methods for analysing the former are directly applicable to the latter, but the calculation of the incoming neutral flux is relatively simple and thus reliable.
In the future, ASCOT4 orbit loss calculations can be further validated using the validated IR and NBI models.

This paper presents a detailed study of a single W7-X experiment program without a plasma.
One NBI source was injecting at full power for \SI{400}{\milli\second} and thus heating up the carbon beam dump, parts of which are monitored with the IR cameras as well as on-line safety measurements.
The central result is a comparison of synthetic and measured camera frames, that are presented as a validation of the modelling (section~\ref{sec:beamFootPrintIr}).
Section~\ref{sec:divergMeas} presents additional spectral measurements of the beam geometry and a test of the camera optical performance is part of the section~\ref{sec:MTF}.
The measurement results are supported by a brief description of the used NBI, IR and spectroscopy hardware in section~\ref{sec:harware} as well as description of the modelling tools and the thermal model in section~\ref{sec:methods}. The Summary, Discussion and Future Work section~\ref{sec:summary} discusses i.a. the level of agreement between measurements and modelling as well as the larger than expected divergence of the beams.


\section{Hardware}
\label{sec:harware}
This section provides a brief overview of the studied NBI injector and the used measurement apparatus. First the NBI injectors and some of the associated instrumentation is described, then the beam dump and how it is instrumented.



In the 2018 experiments, the first W7-X NBI box (NI21) was commissioned with two sources each injecting approximately \SI{1.7}{\mega\watt} of up to \SI{55}{\kilo\electronvolt} hydrogen atoms~\cite{selectionOfW7xStartUpBeamSources,CurrentStatusOfW7Xnbi2013}. The second box with further two sources will be ready for the next experiments in 2021+. Both boxes have the option to install two more sources each and to produce deuterium beams.

In the NBI sources, H$^{+}$, H$_{2}^{+}$, and H$_{3}^{+}$ ions are extracted from a plasma by means of 3 grids, each with 774 apertures, where the middle grid prevents back-acceleration of electrons~\cite{Kraus1993}. In each source, the two grid halves are tilted vertically by 0.87\textdegree\ to focus the beam halves. In addition, the holes in the plasma facing grid halves have an offset that increases linearly with distance from the grid half centre, to induce an inward steering of the outer beamlets that effectively focuses the beam halves. To correctly model the beam profile and emission, the single beamlets need to be taken into account. 

Directly after the source, the extracted ions interact with H$_{2}$ background gas in a tapered pipe, the neutraliser, leading to dissociation of the molecular ions, resulting in three energy components, and neutralisation.
In addition, background gas interaction leads to excitation and light emission. 
Balmer-\textalpha{} (H-\textalpha{}) emission is collected along a vertical line of sight to diagnose the beam. 
The optics is a \SI{135}{mm} f2.0 objective that couples the light to an optical fibre for transmission to an \emph{Ocean Optics HR4000} spectrometer measuring the wavelength range from \SIrange{625}{678}{\nano\meter} using \SI{100}{\milli\second} integration time.
After the neutraliser, a magnet bends the remaining ions into an ion dump within the NBI box.
In W7-X, the neutral beam then leaves the NBI box and passes through a narrow carbon covered beam duct, that scrapes off the edges of the beam. 
The NBI intersects the plasma in a near perpendicular angle resulting birth pitch ($v_\parallel/v$) distributions centred around 0.3 and 0.5 depending on the used NBI source. Finally, the neutrals that pass through the plasma without ionising hit the beam-dump at the far side of the plasma.



The beam dump is made of carbon tiles bolted onto CuCrZr heat sinks with a Sigraflex layer in between. The sinks are brazed on stainless steel water cooling pipes~\cite{w7xPFCthermomechanics}. 
Overheating of the structure may lead into e.g. failures in the brazing or phase change of the CuCrZr. Thus, the surface temperature of certain tiles in the beam dump is monitored with infrared cameras and pyrometers as well as the bulk copper temperature with thermocouples~\cite{W7XnbiDumpThermocoupleSafetyInterlock}.   The measurement geometry is drafted in figure~\ref{fig:bbnbiVsIonOpts}(f). The thermocouples are too slow reacting for the present study, because they are mounted on the backside of the CuCrZr heat sink.

The NBI beam dump is partly within the wide angle view of the infrared (LWIR) cameras~\cite{IRimagingSys4wallProtectionInW7X}. Each pixel of the camera is a resistive micro-bolometer that measures a temperature rise due to impinging radiation 
in the wavelength range \SI{8}{\micro\meter}--\SI{10}{\micro\meter}. 
The surface temperature is evaluated from tabulated resistivities after the experiment program is over using a calibration look-up table. 
The calibration is Planck law based and was performed with a reference black body radiator, while incorporating the complete optical path, including the vacuum window.

The beam dump surface temperature is continuously monitored by the dedicated heat shield thermography (HST) system (section 2.4.1 of~\cite{IRspectrometryAUG}), that terminates the NBI heating if the surface temperature exceeds a set limit.
The system consists of Keller PA41 quotient pyrometers operating at \SI{950}{\nano\meter} and \SI{1050}{\nano\meter}. However, in the studied program the pyrometer was operated only with the \SI{1050}{\nano\meter} channel. The optical path starts at a purpose built optical head, which is a telescope viewing the beam dump through a sapphire window and focuses the light onto a fiber-bundle. The light is then coupled into a single \SI{800}{\micro\meter} fiber for transmission to the pyrometer.


\section{Methods}
\label{sec:methods}
This section describes the computational models to be validated. The first part describes the NBI modelling tools and the second one describes the synthetic IR camera model. Furthermore, an assessment of the optical performance of the IR camera is given.
\subsection{NBI models}
The main modelling tool used in this paper is the Beamlet Based NBI (BBNBI)~\cite{Asunta14_BBNBI_reference} code.
The main application of the code is to prepare initial NBI ion ensembles for ASCOT simulations by calculating the ionisation locations of the neutral beam particles in the plasma.
In the studied W7-X program, there was no plasma, and BBNBI is used solely to study what is normally called shine-through, i.e. the heat-flux impinging the far wall behind the plasma.

BBNBI uses a fine-grained description of the beam geometry: each aperture in the source is a separately modelled beamlet with a location, direction and divergence.
The divergence describes how the particle jet spreads as it propagates from the aperture. The  distribution of the beamlet is assumed to be radially symmetric around the beamlet axis and the angular dependency is assumed to be Gaussian in shape. The definition of divergence used in BBNBI and in this paper is the half-width-1/e-maximum angle. 

The extracted power is determined from measurements of the acceleration voltage and extracted current. 
The NBI-neutraliser spectroscopy is used to determine the fractions for the full, half, and one-third energy particles, which are needed to determine the neutralisation efficiency and thus the neutralised beam power. 
In the ion source, H$^{+}$, H$_{2}^{+}$, and H$_{3}^{+}$ are extracted. Collisions with background gas in the neutralizer lead to dissociation, neutralisation, and light emission. The reaction products originating from the separate extracted species move at different speeds, which leads to a different Doppler shift of their emission. The extracted current fractions are determined by comparing the measured emission with the modelled relative intensity for the different components.
The systems of differential equations that describes the evolution of the separate extracted species are numerically integrated as function of target thickness using the cross sections from~\cite{Tabata2000}. Missing reactions are supplemented from~\cite{Berkner1975}. With the reaction fragments per extracted species, the relative emission intensities per energy component are calculated with the Balmer-$\alpha$ excitation cross sections from~\cite{Williams1982, Williams1983}. This allows to convert the measured relative emission intensities to the extracted current fractions with the correction factor approach~\cite{Hemsworth1985}. The measured extracted current is used together with the extracted current fractions and calculated species evolution to arrive at a particle flux per energy component. 
The exact gas density in the neutraliser is not known. 
From experience at other experiments, a neutraliser thickness of \SI{2.5e20}{\per\meter\squared} is assumed. 
This results in a neutralisation efficiency of $\eta$=\SI{53}{\percent}, which is close to the infinite target thickness solution.
The efficiency is in agreement with calorimetric measurements.
Further losses e.g. in the beam duct are expected due to reionisation and beam scraping.

The BBNBI code calculates the shine through wall loads by Monte-Carlo method. It repeatedly randomly chooses a beamlet and the direction from the divergence distribution to generate a marker. Normally the ionisation location of the marker is calculated by integrating the ionisation probability along the neutral trajectory, but in this case only the wall hit location is needed. The code uses a detailed model of the beam duct and beam dump geometry stored in a triangular surface mesh to calculate the wall hit location.
Each marker is associated with a number of quantities, i.a. energy $E$ and weight $w$. The weight depicts the number of injected real particles per second the marker represent. Thus, the heat flux $q$ to a wall triangle $i$ can be calculated by a simple summing over markers $j$: $q_i=\sum_j E_j\cdot w_j/A_i$, where $A_i$ is the triangle surface area.

\begin{figure}
\centering%
\setlength{\unitlength}{\linewidth}
\begin{picture}(1.0,1.2)
\put(0.0,0.70){\includegraphics[trim=10cm 0 25cm 17cm, clip,width=0.44\linewidth]{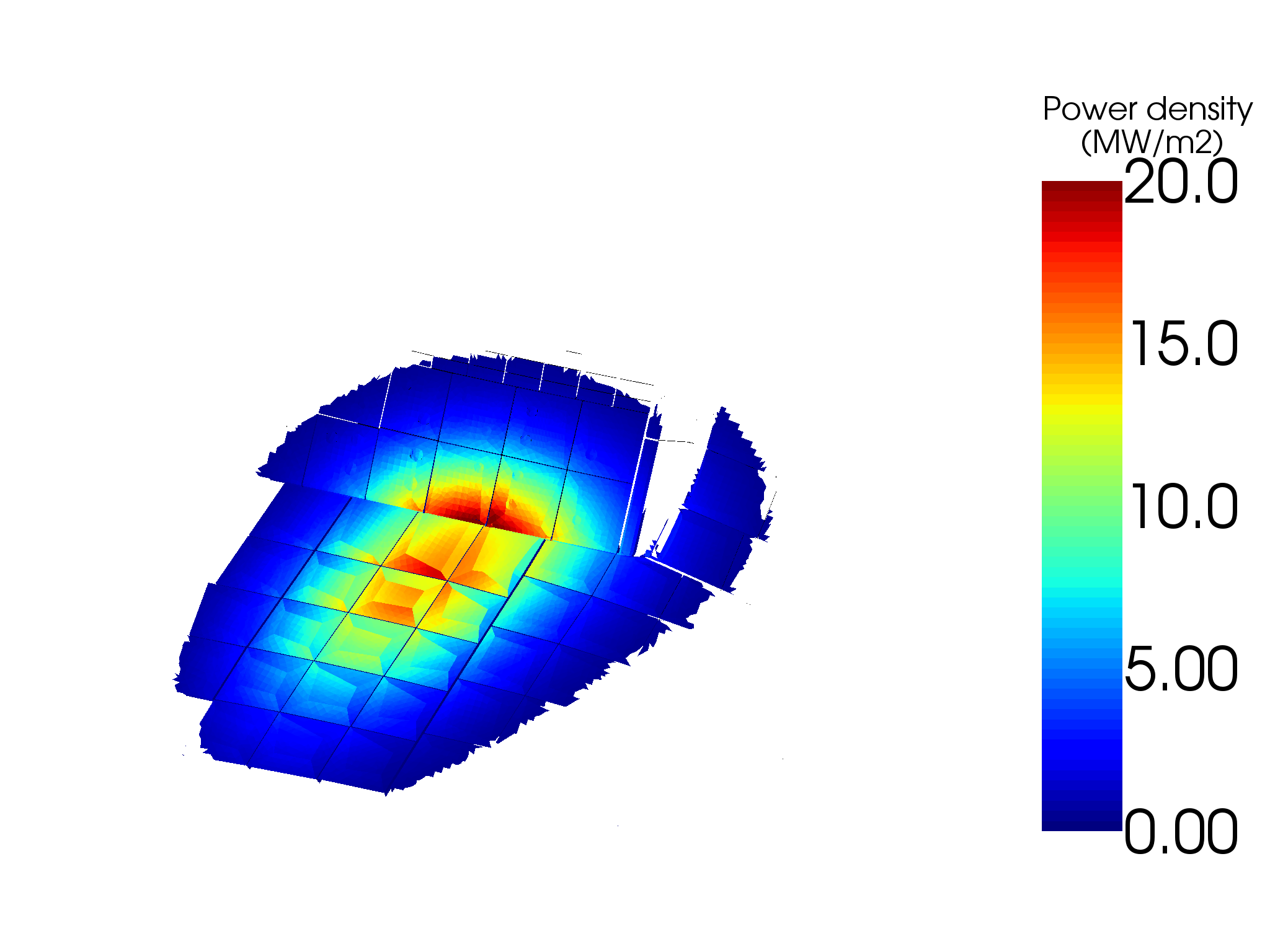}}
\put(0.01,1.10){{\color{black}(a)}}
\put(0.46,0.70){\includegraphics[trim=10cm 0 25cm 17cm, clip,width=0.44\linewidth]{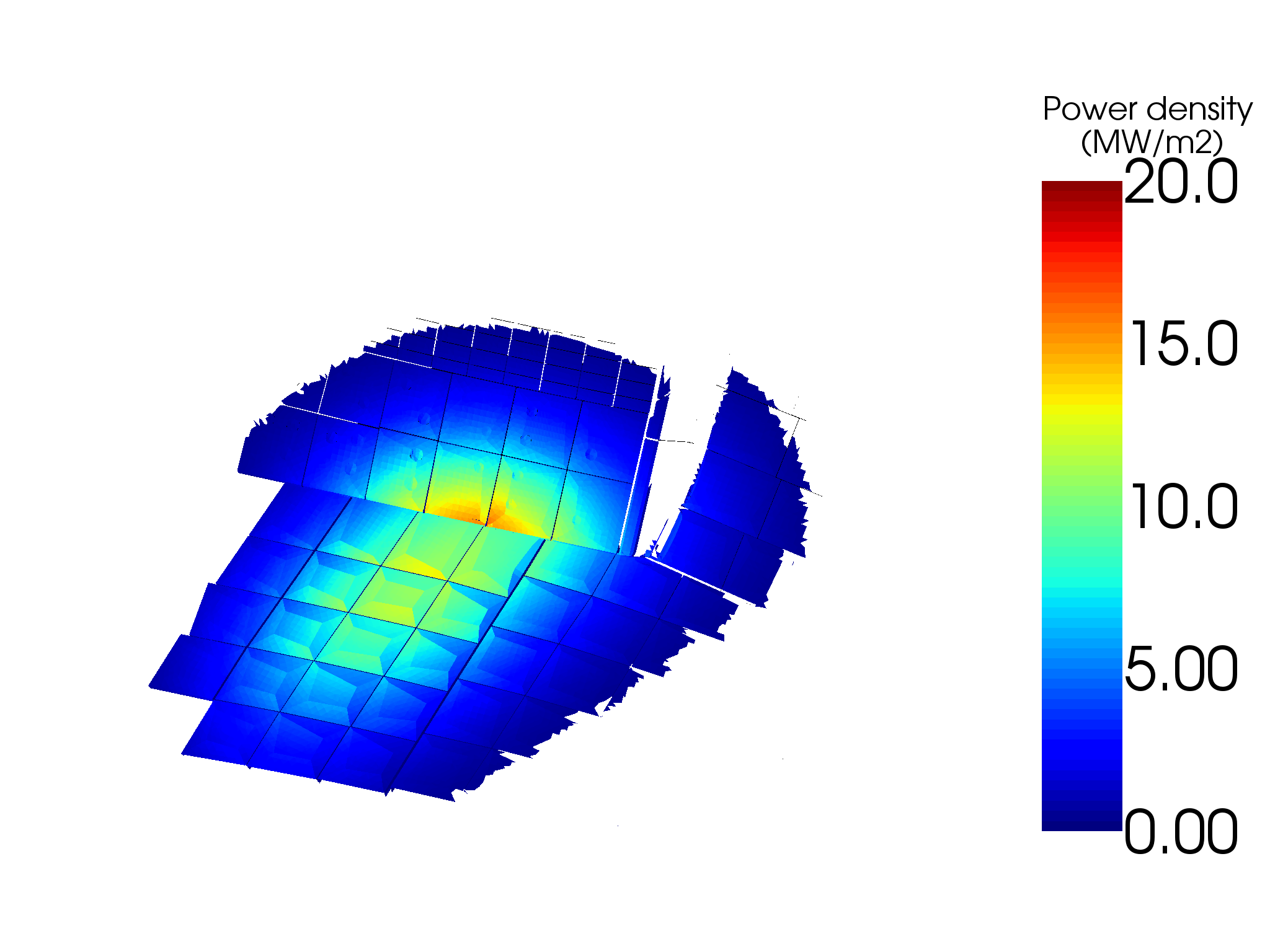}}
\put(0.9,0.80){\includegraphics[trim=55.5cm 0 1.5cm 0cm, clip,width=0.1\linewidth]{{W7-X_NI21_innerWall_Tiles_with_sidest_v2_1.20_deg_Q7_2MW_20.00_MWm2_white}.png}}
\put(0.51,1.10){{\color{black}(b)}}
\put(0.0,0.38){\includegraphics[trim=2cm 5cm 7cm 0cm, clip,width=0.44\linewidth]{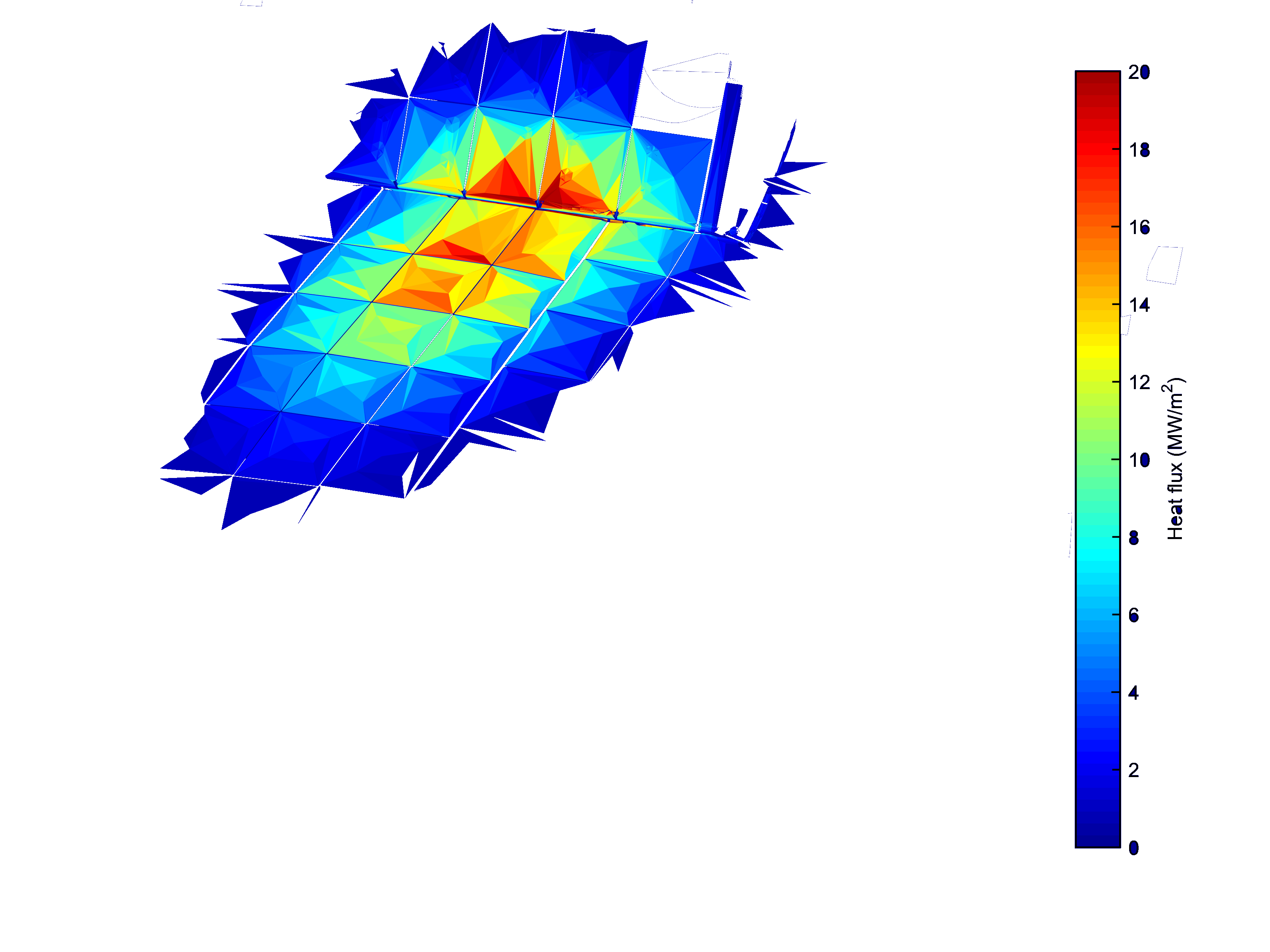}}
\put(0.01,0.78){{\color{black}(c)}}%
\put(0.46,0.38){\includegraphics[trim=2cm 5cm 7cm 0cm, clip,width=0.44\linewidth]{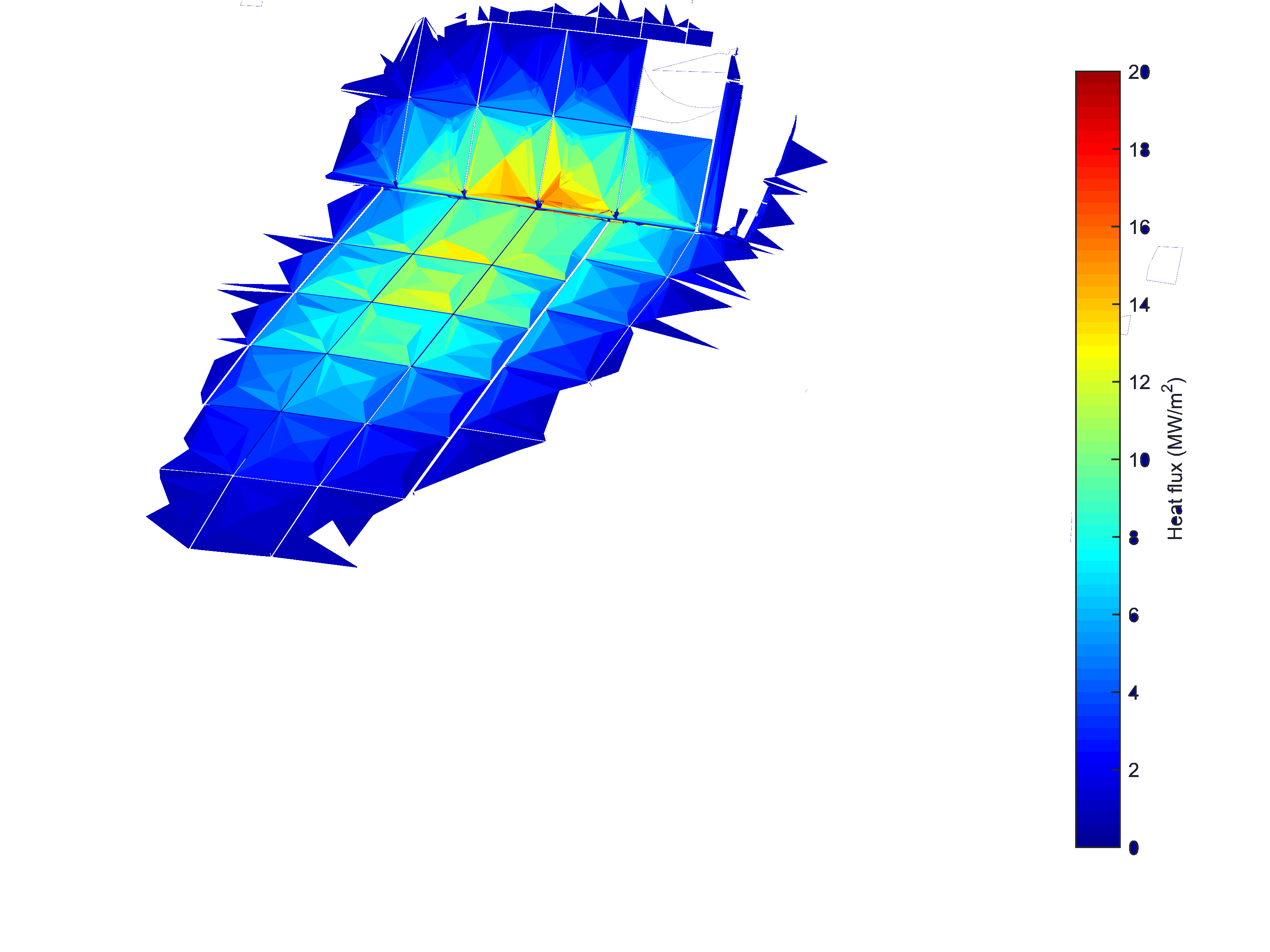}}
\put(0.51,0.78){{\color{black}(d)}}%
\put(0.0,0.0){\includegraphics[trim=2cm 5cm 7cm 0cm, clip,width=0.44\linewidth]{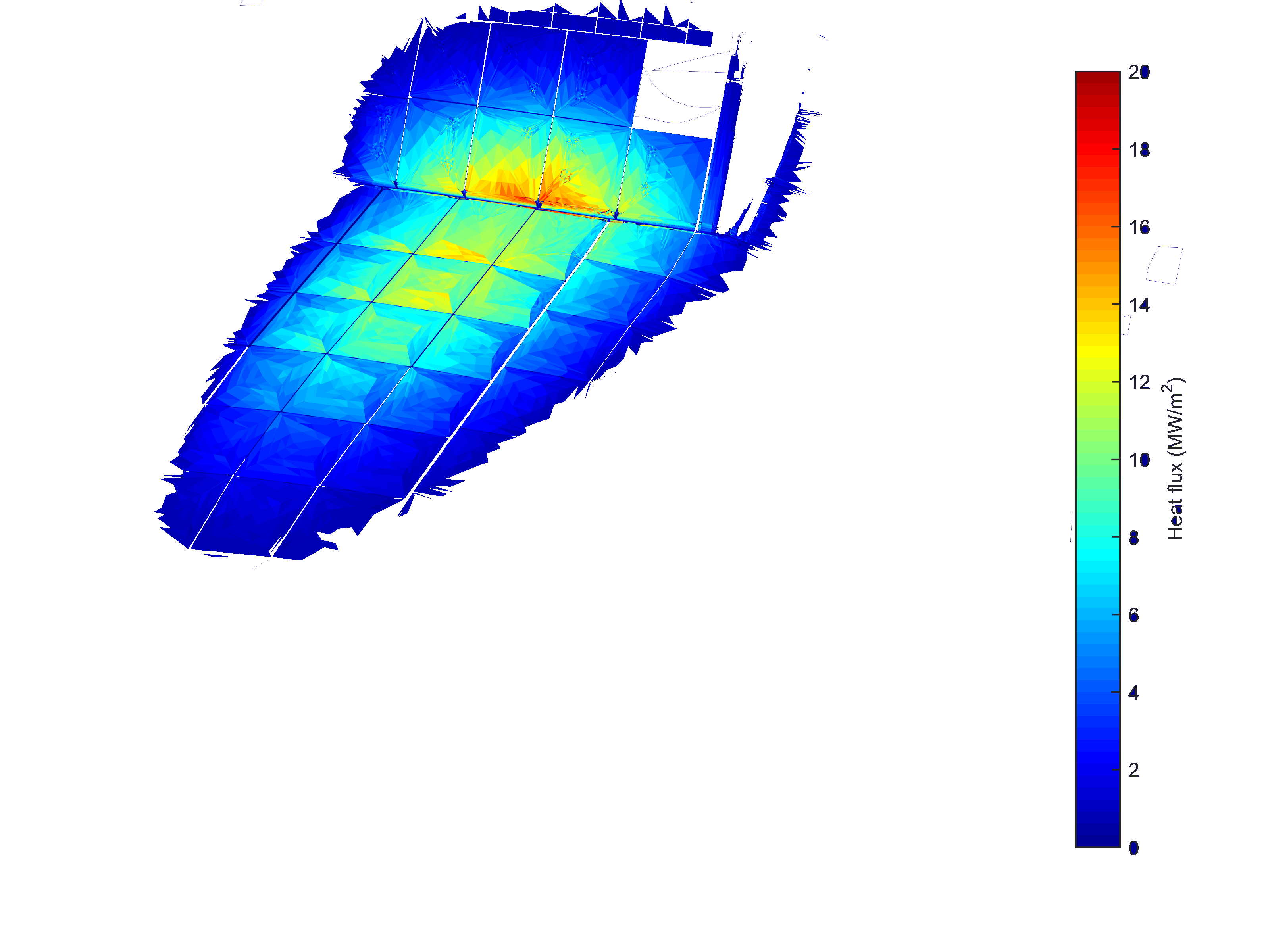}}%
\put(0.01,0.4){{\color{black}(e)}}
\put(0.46,0.0){\includegraphics[trim=8.8cm 22.0cm 30.8cm 0cm, clip,width=0.44\linewidth]{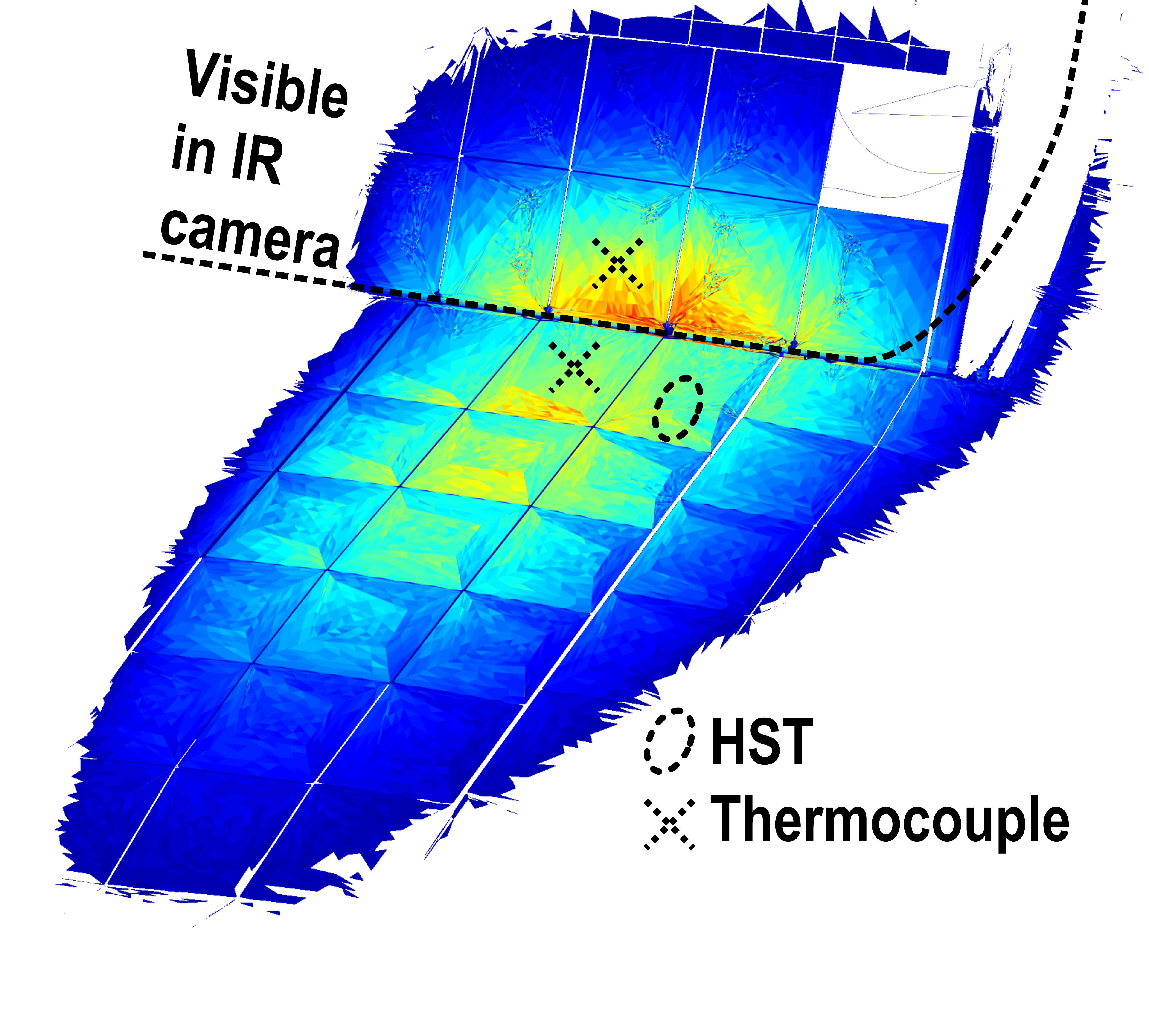}}%
\put(0.51,0.4){{\color{black}(f)}}
\end{picture}%
    \caption{Verification of BBNBI vs. a model with analytical divergence distribution~\cite{NBIAUGtransbeamlinelosses}. Panels (a) and (b) present the heat load due to NBI source 7 injecting \SI{2}{\mega\watt} of neutralized power. The divergences are \SI{1.0}{\degree} and \SI{1.2}{\degree} in (a) and (b), respectively. Panels (c) and (d) are results with the same settings using BBNBI. Panels (e) and (f) show how results from (d) change when each wall triangle is split into 9 and 25 similar subtriangles, respectively. Panel (f) also indicates the tiles instrumented with thermocouples, the field of view of the heat shield thermography (HST) system and approximately the part of the beam dump visible to the IR camera. }
    \label{fig:bbnbiVsIonOpts}
\end{figure}

Where BBNBI calculates the power density employing Monte-Carlo principles and generates markers launched at random directions, another, independent, implementation~\cite{NBIAUGtransbeamlinelosses} of the same geometry rather sums the contribution of each beamlet for each triangle and hence avoids the Monte-Carlo noise. Comparisons of these two implementations are presented in figure~\ref{fig:bbnbiVsIonOpts} and act as a verification of the models.

\subsection{Scene model and Synthetic IR model}

The synthetic camera model is build on the so-called scene model.
For each infrared camera, a pinhole camera model has been created through spatial calibration~\cite{TowardsImageProcW7XspatCalib2thermalEvents} by taking into account the lens distortion of the camera. By using the camera model and a simplified CAD (Computer Aided Design) model of the observed PFCs, the projection and distortion model of the camera is reconstructed and thus the so-called scene model~\cite{W7XnRTImageDiagnosticPFCprot} is created.
Each scene model stores, for each pixel, information about the properties of the target elements, in terms of: spatial coordinates, PFC index, emissivity of the material, angle between the line of sight and the normal to the surface, and distance from the camera. 

\begin{figure}
    \centering
    \includegraphics[width=0.5\linewidth]{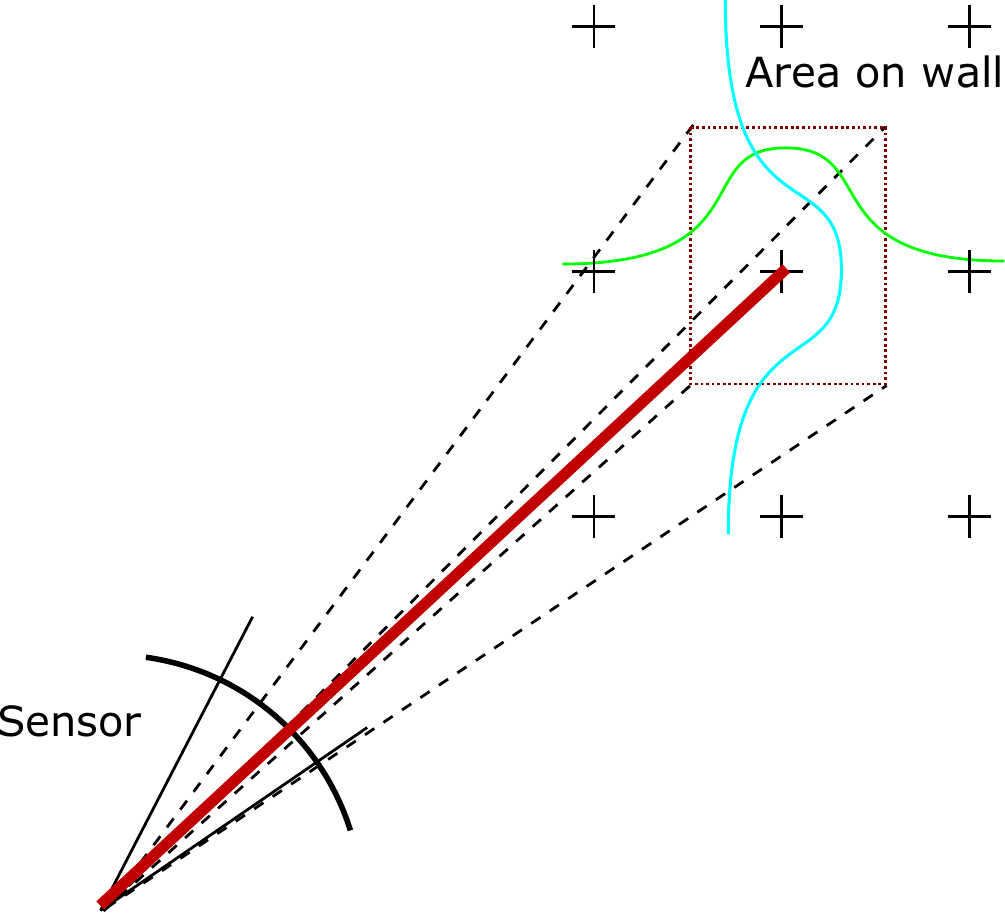}
    \caption{The concept of camera smoothing model. The red line shows the centre axis of a pixel's line of sight, that intersects the wall at the $+$-sign. The eight neighbouring pixels are centred at the other $+$-signs. Ideally, the pixel would be uniformly sensitive to the thermal radiation coming from the wall inside the rectangle marked with dashed lines. In reality, the pixel is likely to be more sensitive to radiation near the pixel's axis and there is some cross-talk between pixels. The green and magenta curves describe hypothetical distributions in the horizontal and vertical directions, respectively.}
    \label{fig:smoothing}
\end{figure}

Ideally, each pixel in the thermal camera would be uniformly sensitive to thermal radiation emitted in a rectangle shaped section of the wall.
In reality, there is some degree of cross-talk (blurring) between pixels and the sensor is likely to be more sensitive near the centre of the pixels field of view than at the edges.
This is illustrated in figure~\ref{fig:smoothing}.
Such effects are important when imaging sub-pixel sized features on the wall.
The effects are accounted for when generating synthetic IR images from BBNBI generated wall loads/wall surface temperatures.
In practise, each synthetic pixel is supersampled by ray-tracing 300 rays from the IR camera aperture towards the wall.
The rays are normally distributed around the pixel axis (as defined in the scene model) with blurring factor $\sigma$ being the standard deviation normalised to the half-distance between the pixels.

The supersampling, or averaging, of the surface temperature with the 300 rays is a non-linear operation due to the spectral emissivity being a highly non-linear function of temperature.
To simplify the calculation, a number of quantities are assumed to be constants for all the rays in the average, and are grouped  into a single constants $\chi$, that cancels out in the calculation. 
The quantities include: 
\begin{enumerate*}
    \item number of photons: $hc/\lambda$;
    \item surface area of the point $j$: $A_j$;
    \item the solid angle of the pixel: $\Omega$;
    \item the size of the wavelength window $\Delta\lambda$;
    \item the emissivity $\varepsilon$ of material (black body to grey body).
\end{enumerate*}
%
The spectral emissivity $\epsilon$ comes directly from the Planck's law, as a function of temperature $T$:
\begin{equation}
    \epsilon(T) =\chi\sum_i P(\lambda_i,T) = \chi\sum_i \frac{2 h c^2  }{\lambda_i^{5}\exp{\frac{h c }{\lambda_i k_\textrm{B}  T}-1}} 
\end{equation}
where the sum is over the two wavelengths at the boundaries of the sensitivity of the detector: $\lambda_i=\{\textrm{\SI{8}{\micro\meter}}, \textrm{\SI{10}{\micro\meter}} \}$.  The constants are the usual: speed of light $c$, the Planck constant $h$ and the Boltzmann constant $k_\textrm{B}$.
The actual average over ray-wall point intersections $j$ is given by:
\begin{equation}
E = \frac{1}{n}\sum_{j=1}^n \epsilon(T_0+\Delta T_j)
\end{equation}
The initial temperature is assumed to be uniform $T_0 =  \SI{100}{\celsius}$, and the temperature change between two IR camera frames is $\Delta T$.
The final averaged temperature is given by the inverse of the spectral emissivity function: $T=\epsilon^{-1}(E)$, which is realised as an inverse interpolation of a look-up-table calculated with $\epsilon$. 
    
\label{sec:MTF}
The optical performance of the IR cameras were measured in the real measurement geometry recreated in the laboratory. The test object was a hot source at \SI{300}{\celsius} with a slitted screen. The screen had two sets of vertical slits with \SI{6}{\milli\meter}
and \SI{12}{\milli\meter} 
width which created an interleaved stripe pattern on the IR image. The limited sharpness of the imaging system can be characterised by the modulation transfer function (MTF) which can be derived from these pattern in the images. 
In order to optimise the synthetic IR model, the measured profiles of the stripe pattern were compared to profiles in a synthetic image of a similar setup. Thus, a realistic blurring factor $\sigma$ is estimated and the best fit is found with $1.5<\sigma<2.0$.



The temperature change in this work is due to the heating up of the plasma facing components due to a short, strong pulse of heat in the form of a flux of fast neutrals.
The heat has insufficient time to conduct to the back side of PFCs during the NBI pulse.
This is modelled with the simple and well known model, where a flat surface divides the space in two.
One side is filled with uniform, isotropic heat conducting material and the initial temperature is uniform $T_0$. 
At time $t=0$, a constant uniform heat flux $q$ is imposed through the surface.
The solution~\cite{CarlslawConductionOfHeatInSolids} to the heat equation is
\begin{equation}
\Delta T(t,x)=\frac{qx}{\lambda}\left(\frac{1}{\sqrt{\pi}F}\exp(-F^2)-\mathrm{erfc}(F) \right), \label{eq:semiInfinitePenetration}
\end{equation}
%
The constants used in this study for the studied carbon components were measured using laser flash analysis and are as following: heat conductivity $\lambda$=\SI{61}{\watt\per\kelvin\per\meter}; density $\rho$=\SI{1814}{\kg\per\meter\cubed}; and specific heat $c_p$=\SI{1734}{\joule\per\kg\per\kelvin}. 
The change of surface temperature is given by the formula~\cite{CarlslawConductionOfHeatInSolids}
\begin{equation}
\Delta T(t,0) = q\frac{2}{\sqrt{\pi\,\lambda\rho c_p}}\sqrt{t}.  \label{eq:semiInfiniteSurface}
\end{equation}

Radiative cooling of the surface is insignificant compared to the heat flux from the NBI injected neutrals:
according to the Stefan-Boltzmann law, a \SI{1000}{\celsius} surface emits only approximately \SI{150}{\kilo\watt\per\meter\squared}, assuming a perfect emitter (a black body).

\section{Measurements and modelling results}

The main goal of this work is to validate the synthetic IR and NBI models.
Before this, the NBI divergence 
is assessed from measurements. 
After this, the actual validation is performed and, as a final topic, the independent temperature measurement from the HST system is compared to the modelling.


\subsection{Divergence measurement by Beam Emission Spectroscopy in the NBI neutraliser}
\label{sec:divergMeas}


Emission spectra contain information about the angular distribution of the beam, because of the Doppler shifts due to the velocity component parallel to the spectroscopic line of sight. Emission spectra measured in the neutraliser are used to deduce the distribution by comparing them with calculated spectra. 

Calculated spectra are generated from the wavelength and intensity contributions of each beamlet to each subvolume in the observation cone. Spectra are calculated for a fixed beamlet steering factor of \SI{3}{\degree\per\milli\meter}, which describes the strength of the beamlet offset steering leading to focusing of the whole beam, and several divergences, which describe the angular distribution of the particles within a beamlet. After convolution with the instrument function, measured with a neon calibration lamp, the calculated spectrum is scaled and allowed a small shift to minimize the sum of square difference between synthetic spectrum and measurement. To estimate the relative goodness of fit, the coefficient of determination $R^{2}$ is calculated.

\begin{figure}
    \setlength{\unitlength}{\linewidth}
\begin{picture}(1.0,0.34)
\put(0.01,0.33){{\color{black}(a)}}
\put(0.51,0.33){{\color{black}(b)}}
\put(0.5,0.0){\includegraphics[width=0.49\linewidth]{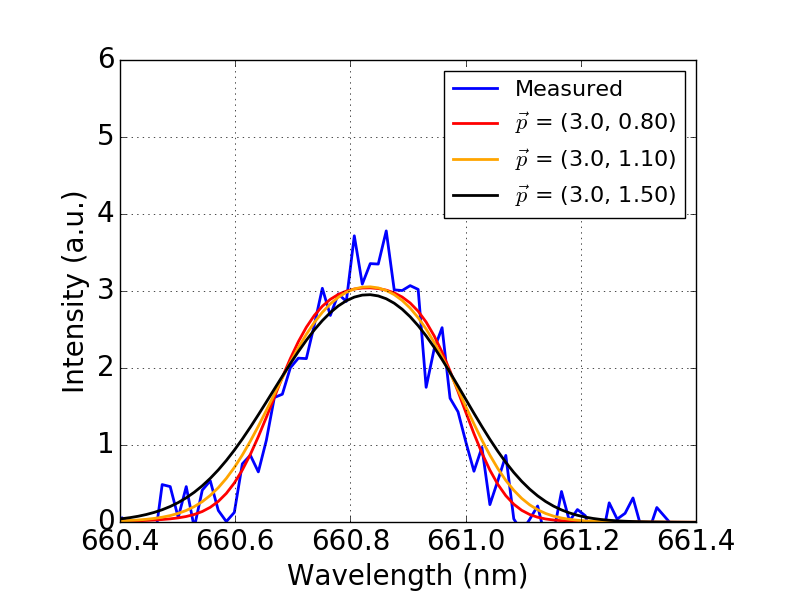}}%
\put(0.0,0.0){\includegraphics[width=0.49\linewidth]{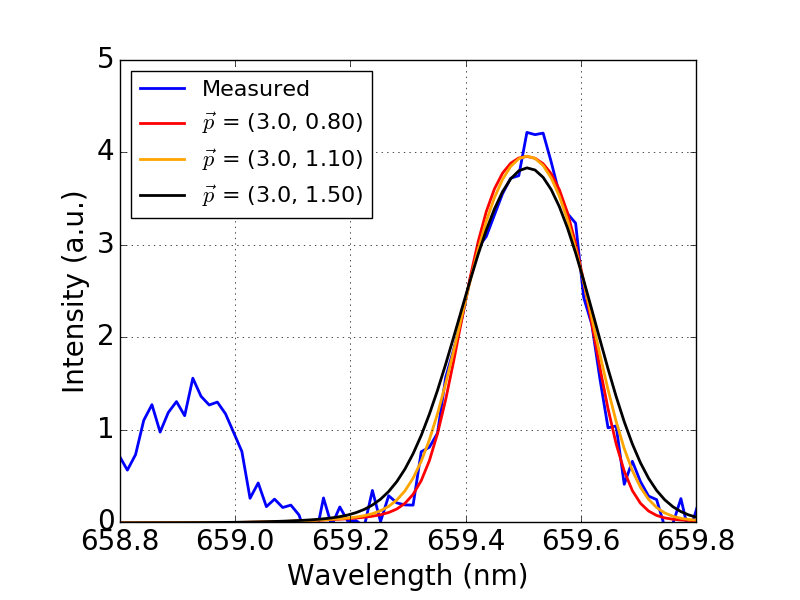}}%
\end{picture}%
    \caption{Measured beam emission spectra with calculated spectra for the full energy (a) and half energy (b) component. Calculated spectra are shown for several divergences. The differences between measurement and the different synthetic spectra are relatively small, leading to relatively large errorbars. At a fixed beamlet offset steering of \SI{3.0}{\degree\per\milli\meter}, which is output from ion-optics calculations, the most likely divergence is \SI{0.8}{\degree} for the main energy component, and \SI{1.0}{\degree} for the half energy component.
        \label{fig:divergenceMeasurement}}
\end{figure}

Beam emission spectra were measured for an acceleration potential of \SI{55}{\kilo\volt}, at an extracted power of \SI{5}{\mega\watt}.
The spectra indicated that the beam consisted of \SI{25}{\percent} full-energy particles, \SI{63}{\percent} of half-energy particles and \SI{12}{\percent} of 1/3-energy particles. 
Figure~\ref{fig:divergenceMeasurement} shows the full and half energy component of the spectrum with several synthetic spectra.
The different divergences provide a similar match to the spectrum. At a fixed beamlet offset steering of \SI{3.0}{\degree\per\milli\meter}, which is output of ion-optics calculations, the most likely divergence is \SI{0.8}{\degree} for the main energy component, and \SI{1.0}{\degree} for the half energy component. When one chooses as error estimates the divergence at which the $R^{2}$ is 0.995 with respect to the best fit, the uncertainties are approximately 0.3~degrees in both directions.
The W7-X divergences are close to the values determined in the same way on similar sources at ASDEX Upgrade~\cite{NBIAUGtransbeamlinelosses}.

\subsection{IR measurements of the beam dump}
\label{sec:beamFootPrintIr}

A direct comparison between IR camera measurements and synthetic measurements is presented in figure~\ref{fig:IRmeasurements}.
The analysis compares the temperature change during the first \SI{250}{\milli\second} of the NBI heating in the W7-X program 20180918.19. To streamline the comparison, only the NBI source 7 was energised.
The synthetic image was generated for multiple values for the beam divergence as well as for camera blurring factor $\sigma$ describing how well the IR camera was focused and for two values of the neutralisation efficiency $\eta$. 
The best fit to measurements occur near the values obtained in sections~\ref{sec:divergMeas} and~\ref{sec:MTF}: $\sigma=1.0\cdots2.0$, $\eta$=\SI{41}{\percent} and divergence of  \SI{1.2}{\degree} or slightly more. The measured peak values ($\sim$\SI{500}{\kelvin}) and overall shape matches very well. Similar details are visible in both, synthetic and measured, frames and there is an offset of only a few pixels. However, changes in the divergence and neutralisation efficiency result in similar changes of peak heat load and no single parameter-set is found that is outstandingly superior.  

\newcommand{\IRscale}{0.5}%
\newlength{\IRtrimRight}\setlength{\IRtrimRight}{38mm}%
\newlength{\IRtrimLeft}\setlength{\IRtrimLeft}{10mm}%
\newlength{\IRtrimLeftNoLabel}\setlength{\IRtrimLeftNoLabel}{19.5mm}%
\newlength{\IRtrimTop}\setlength{\IRtrimTop}{14mm}%
\newlength{\IRtrimBot}\setlength{\IRtrimBot}{0mm}%
\newlength{\IRtrimBotNoLabel}\setlength{\IRtrimBotNoLabel}{12.5mm}%
\newlength{\IRheightT}\settoheight{\IRheightT}{\includegraphics[scale=\IRscale,trim=\IRtrimLeftNoLabel{} \IRtrimBotNoLabel{} \IRtrimRight{}     \IRtrimTop{},clip]{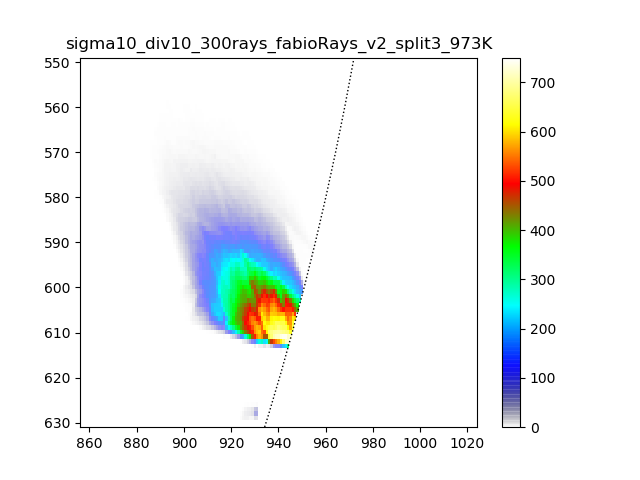}}%

\begin{figure}
\setlength{\unitlength}{1cm}%
%
%
\centering%
\begin{picture}(6.0,5.2)
\put(0.0,0.0){\includegraphics[scale=\IRscale,trim=\IRtrimLeft{} \IRtrimBot{} 0.0 \IRtrimTop{},clip]{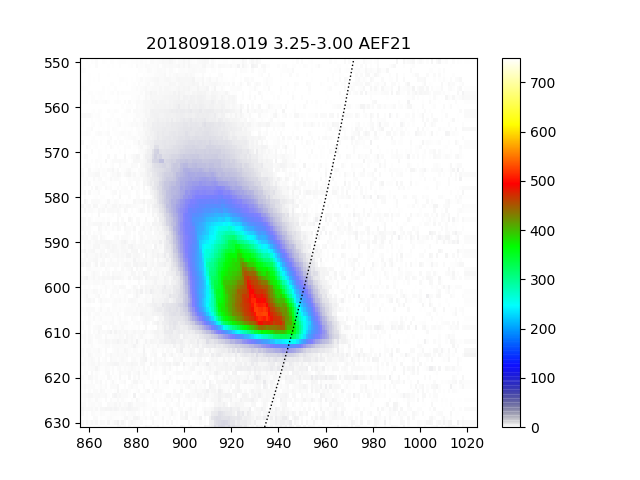}}
\put(0.7,4.9){\textbf{Meas.}}
\put(3.1,4.9){20180918.019}
\put(3.4,4.4){3.25s - 3.00s}
\put(4.3,3.9){AEF21}
\end{picture}\\%
\begin{picture}(18.0,15.)
    \put(0.0,10.4){\includegraphics[scale=\IRscale,trim=\IRtrimLeft{} \IRtrimBotNoLabel{} \IRtrimRight{} \IRtrimTop{},clip]{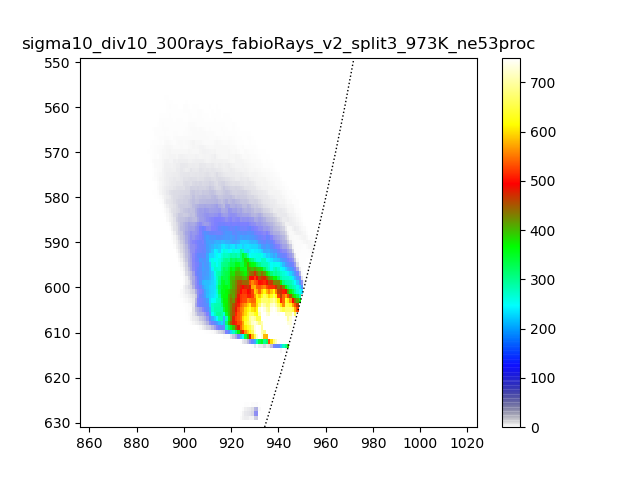}}
    \put(5.6,10.4){\includegraphics[scale=\IRscale,trim=\IRtrimLeftNoLabel{} \IRtrimBotNoLabel{} \IRtrimRight{} \IRtrimTop{},clip]{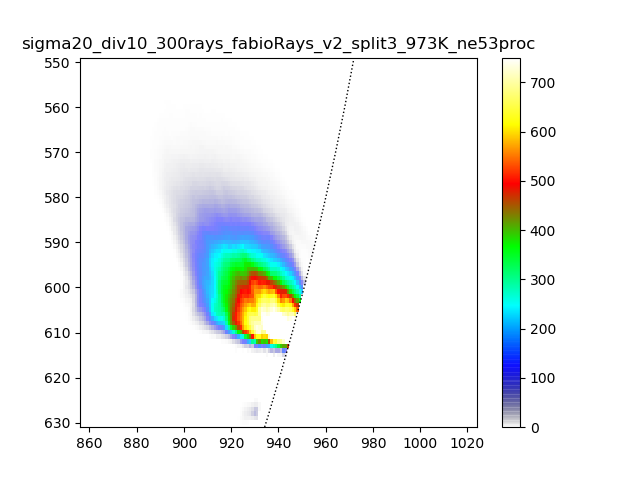}}
    \put(10.75,10.4){\includegraphics[scale=\IRscale,trim=\IRtrimLeftNoLabel{} \IRtrimBotNoLabel{} \IRtrimRight{} \IRtrimTop{},clip]{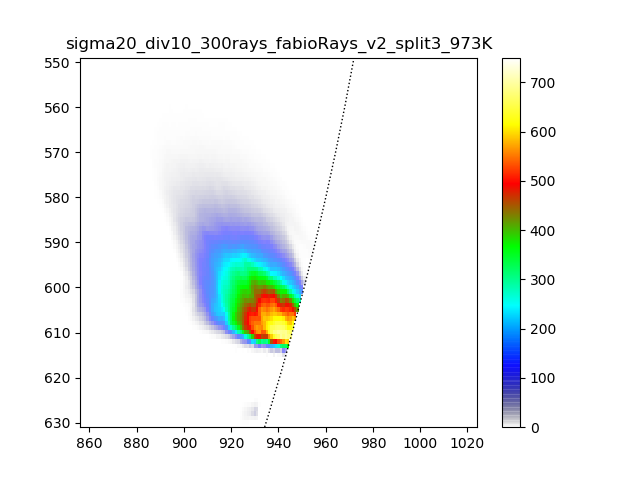}}
    \put(0.7,14.7){\textbf{Synth.}}
    \put(4.1,14.7){$\sigma$=1.0}
    \put(4.0,14.2){div=\SI{1.0}{\degree}}
    \put(4.2,13.7){$\eta$=\SI{53}{\percent}}
    \put(9.3,14.7){$\sigma$=2.0}
    \put(9.2,14.2){div=\SI{1.0}{\degree}}
    \put(9.4,13.7){$\eta$=\SI{53}{\percent}}
    \put(14.4,14.7){$\sigma$=2.0}
    \put(14.3,14.2){div=\SI{1.0}{\degree}}
    \put(14.5,13.7){$\eta$=\SI{41}{\percent}}
    \put(0.0,5.5){\includegraphics[scale=\IRscale,trim=\IRtrimLeft{} \IRtrimBotNoLabel{} \IRtrimRight{} \IRtrimTop{},clip]{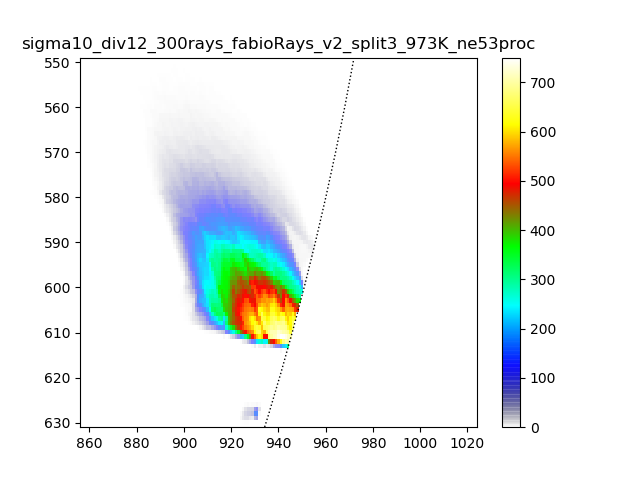}}%
    \put(5.6,5.5){\includegraphics[scale=\IRscale,trim=\IRtrimLeftNoLabel{} \IRtrimBotNoLabel{} \IRtrimRight{} \IRtrimTop{},clip]{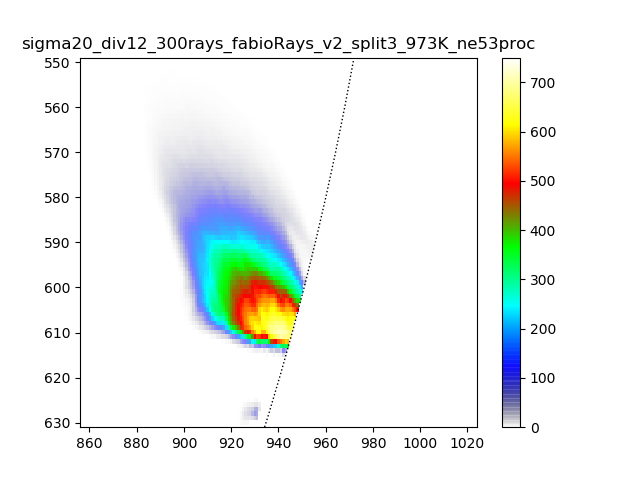}}%
    \put(10.75,5.5){\includegraphics[scale=\IRscale,trim=\IRtrimLeftNoLabel{} \IRtrimBotNoLabel{} \IRtrimRight{} \IRtrimTop{},clip]{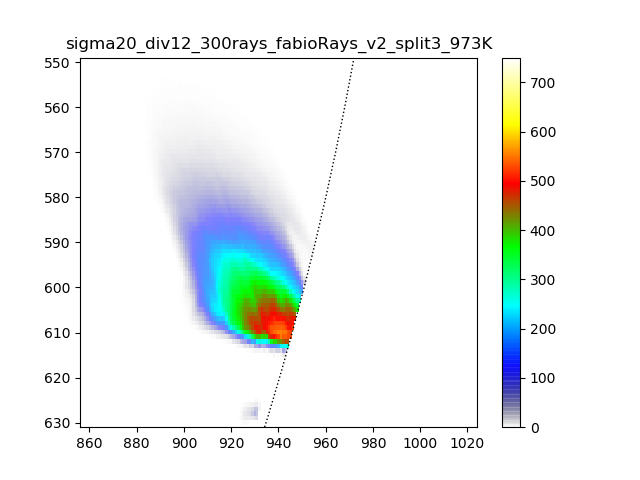}}
    \put(4.1,9.8){$\sigma$=1.0}
    \put(4.0,9.3){div=\SI{1.2}{\degree}}
    \put(4.2,8.8){$\eta$=\SI{53}{\percent}}
    \put(9.3,9.8){$\sigma$=2.0}
    \put(9.2,9.3){div=\SI{1.2}{\degree}}
    \put(9.4,8.8){$\eta$=\SI{53}{\percent}}
    \put(14.4,9.8){$\sigma$=2.0}
    \put(14.3,9.3){div=\SI{1.2}{\degree}}
    \put(14.5,8.8){$\eta$=\SI{41}{\percent}}
    \put(0.0,0.0){\includegraphics[scale=\IRscale,trim=\IRtrimLeft{} \IRtrimBot{} \IRtrimRight{} \IRtrimTop{},clip]{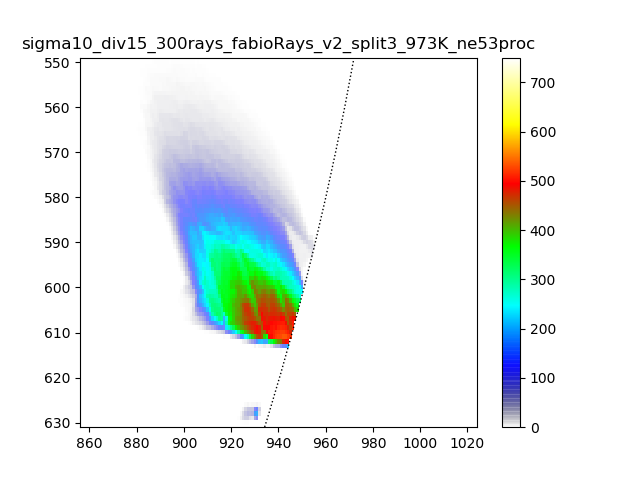}}%
    \put(5.6,0.0){\includegraphics[scale=\IRscale,trim=\IRtrimLeftNoLabel{} \IRtrimBot{} \IRtrimRight{} \IRtrimTop{},clip]{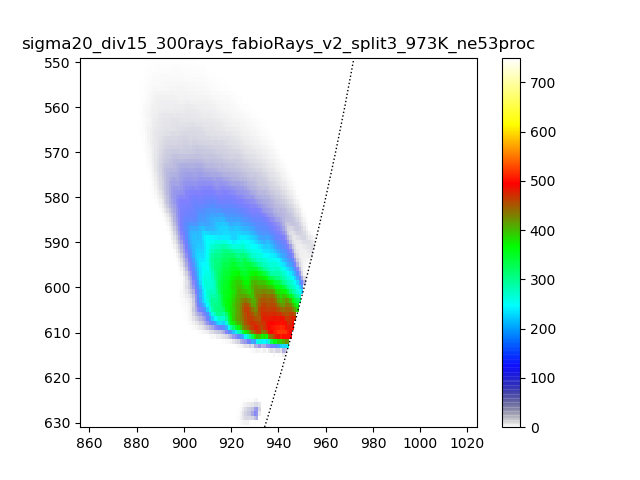}}%
    \put(10.75,0.0){\includegraphics[scale=\IRscale,trim=\IRtrimLeftNoLabel{} \IRtrimBot{} \IRtrimRight{} \IRtrimTop{},clip]{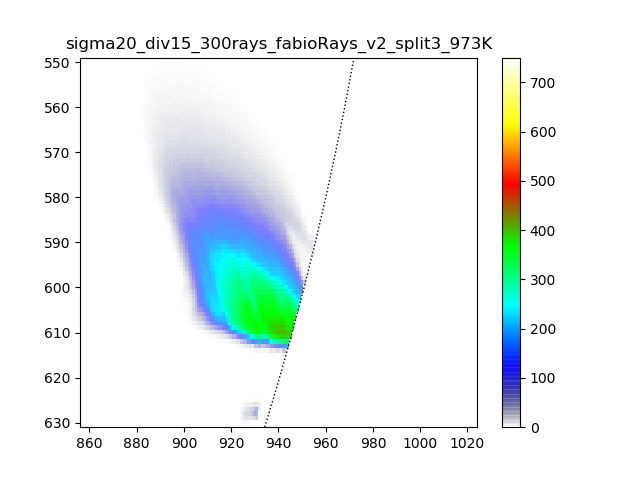}}%
    \put(4.1,4.9){$\sigma$=1.0}
    \put(4.0,4.4){div=\SI{1.5}{\degree}}
    \put(4.2,3.9){$\eta$=\SI{53}{\percent}}
    \put(9.3,4.9){$\sigma$=2.0}
    \put(9.2,4.4){div=\SI{1.5}{\degree}}
    \put(9.4,3.9){$\eta$=\SI{53}{\percent}}
    \put(14.4,4.9){$\sigma$=2.0}
    \put(14.3,4.4){div=\SI{1.5}{\degree}}
    \put(14.5,3.9){$\eta$=\SI{41}{\percent}}
    \end{picture}%
\caption{
Comparison of measured and synthetic IR measurements of the NBI beam dump with various model parameter values. The top panel shows the \emph{measured} temperature difference during the first \SI{250}{\milli\second} of NBI injection using the micro-bolometer IR camera installed in port AEF21. The other panels show \emph{synthetic} IR images with various assumed NBI beamlet divergence angles, IR camera blurring factors $\sigma$ and neutralization efficiencies $\eta$. The dashed line indicates the boundary of well calibrated IR camera pixels. The horizontal and vertical axis values are pixel rows and columns. Colour-scale is temperature change in \si{\kelvin} (identical in all panels).}
    \label{fig:IRmeasurements}
\end{figure}
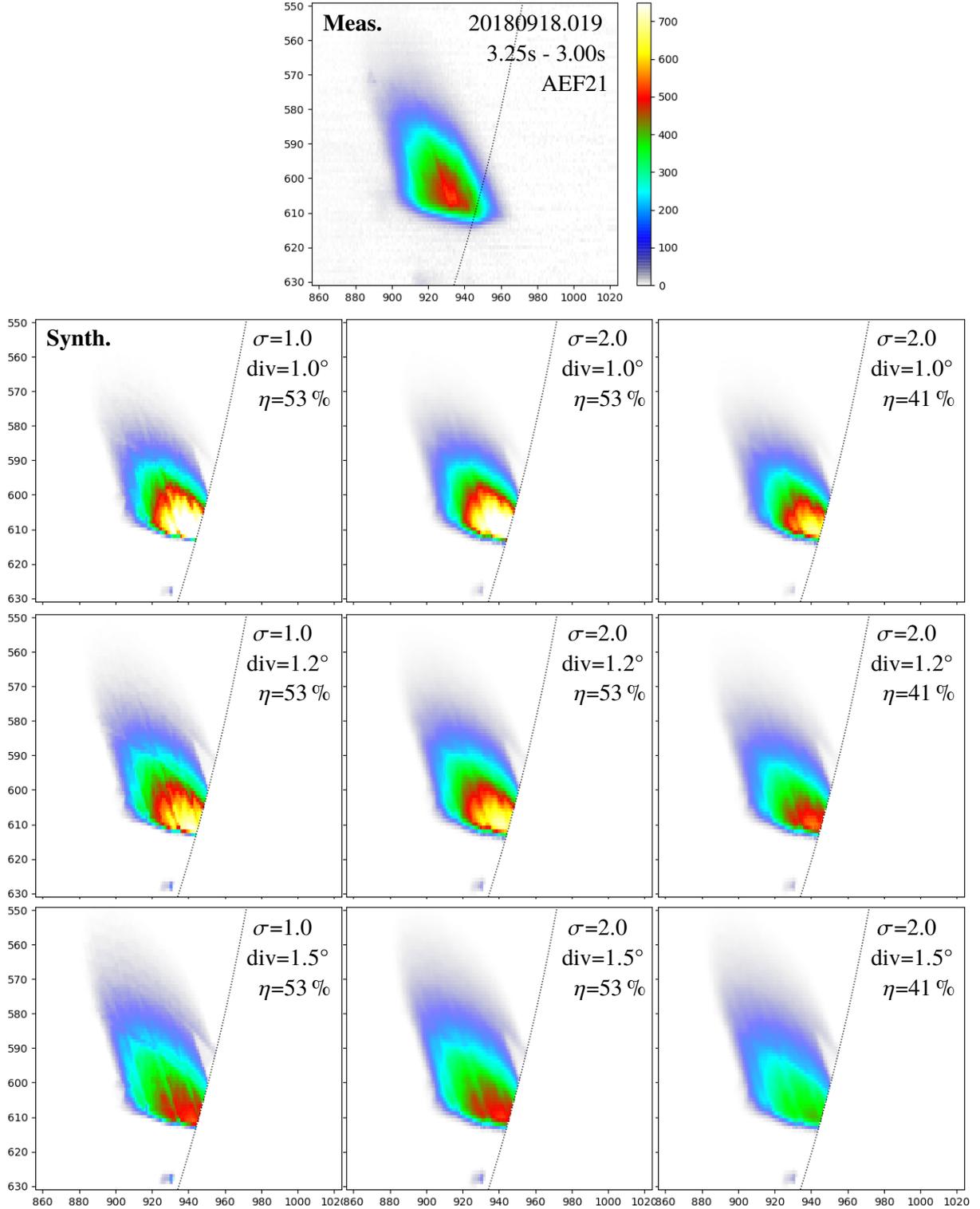

The heat shield thermography (HST) provides an independent IR measurement of the surface temperature during the NBI heating. Direct comparison between the cameras and the HST pyrometer is difficult due to no overlap in field of views and different assumptions of the emissivity $\varepsilon$ in the calibration process. The HST measurement is also reliable only above $\sim$\SI{500}{\celsius}, thus making the initial temperature at the beginning of NBI heating unknown.

\begin{figure}
    \centering
    \includegraphics[width=0.80\linewidth]{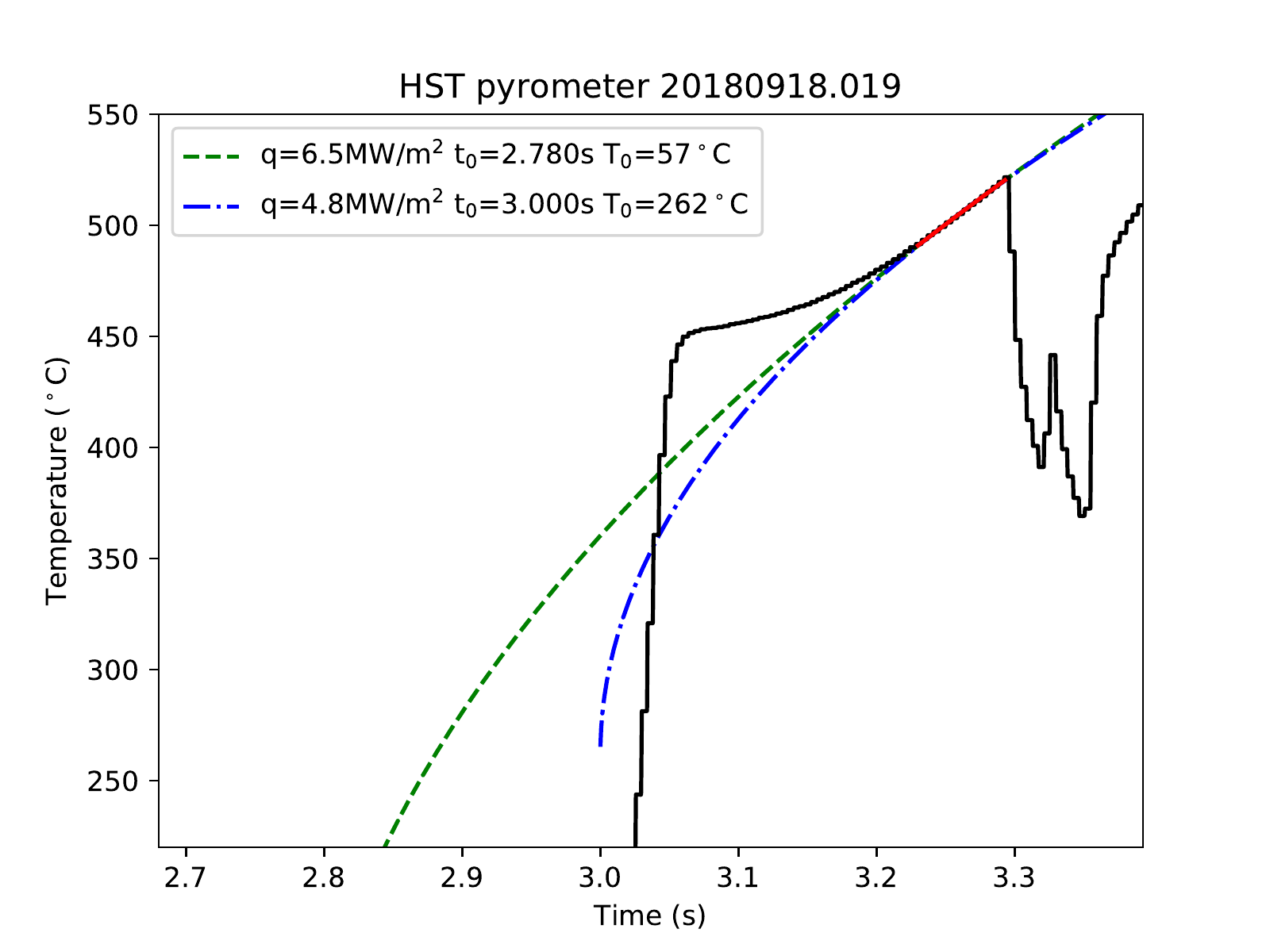}
    \caption{Heat Shield Thermography measurements of beam dump surface temperature. The nominal start of NBI heating was at \SI{3.0}{\second}. The pyrometer is non-linear below $\sim$\SI{500}{\celsius}, so only the last parts of the measurements are useful. The dashed lines indicate fits of eq~\ref{eq:semiInfiniteSurface} to the data shown in red. The green fit has a fixed initial temperature  $T_0$ while the blue one has a fixed initial time $t_0$. The drops in the temperature near \SI{3.3}{\second} are due to arcing in the NBI source. }
    \label{fig:pyro}
\end{figure}

The eq~\ref{eq:semiInfiniteSurface} can be fitted to the temperature data obtained at latter parts of heating, when the temperature is sufficiently high. 
The equation has three free parameters; heat-flux $q$, beam heating starting time $t_0$ and initial temperature $T_0$, which cannot all be fitted simultaneously.
Figure~\ref{fig:pyro} shows the results if one assumes $t_0$=\SI{3.0}{\second} or $T_0$=\SI{57}{\celsius}. The fit gives $T_0$=\SI{262}{\celsius}; $q$=\SI{5.9}{\mega\watt\per\meter\squared} or $t_0$=\SI{2.780}{\second}; $q$=\SI{8.0}{\mega\watt\per\meter\squared}, respectively. The scatter in heat-flux is high, but the values are compatible with modelled heat-fluxes presented in figure~\ref{fig:bbnbiVsIonOpts}, assuming divergence to be \SI{1.2}{\degree}.

The modelling is also in a very rough agreement with a first calorimetric study~\cite{AnnabelleSpanierMaster}: the measurement indicates the beam dump receiving $\sim$\SI{1.2}{\mega\watt} of power while the modelling estimates \SI{1.7}{\mega\watt} to  \SI{2.0}{\mega\watt}.
This should be considered a reasonable result, since the in-vessel water cooling system is sub-optimal for such studies and uncertainties are large.
The discrepancy is mainly attributed to heat loss of the coolant water between the beam dump and the temperature measurement.

\section{Summary, discussion and future work}
\label{sec:summary}

This contribution presents the results of comparing modelled and experimental infrared camera measurements of wall heating due to a short neutral beam injector pulse in the empty Wendelstein 7-X stellarator.
The synthetic infrared image reproduces the measured temperature change both qualitatively (shape) and quantitatively. 
The best matching model parameters (camera blurring factor $\sigma$ and beamlet divergence) are close to values measured with other methods.

The work validates the synthetic IR model and the used NBI model (i.e. the BBNBI code).
The IR camera measurements indicate that either the divergence of NBI injector is larger than expected (\SIrange{1.2}{1.5}{\degree} instead of \SI{1.0}{\degree}) or the neutralisation efficiency is lower or re-ionisation losses higher than expected, which would be in line with experience from other experiments~\cite{NeutralisationIonBeamsJET}. The divergence interval is neither supported nor excluded by the spectroscopic measurements in the NBI neutraliser (div$_\text{E}=$\SI[separate-uncertainty = true]{0.8(3)}{\degree} and div$_\text{E/2}=$\SI[separate-uncertainty = true]{1.0(3)}{\degree}).   The independent pyrometer measurement is compatible with both, the larger divergence or lower efficiency. 
The larger divergence would mainly reduce the efficiency of the NBI system via increased duct-scraping of the beam.

The future work will include using the validated models to analyse NBI orbit loss measurements. 
Further work may include model validation against beam emission spectroscopy in the plasma. 
Possible model developments include asymmetric vertical and horizontal beamlet divergences as well as NBI geometry parameter assessments beyond beamlet divergence. The work could be repeated for other experiments and in the future also for other beam sources.


\acknowledgments

This work was funded partly by Walter Ahlstr\"{o}m foundation. This work has been carried out within the framework of the EUROfusion Consortium and has received funding from the Euratom research and training programme 2014-2018 and 2019-2020 under grant agreement No 633053. The views and opinions expressed herein do not necessarily reflect those of the European Commission.




\bibliographystyle{unsrturl}
\bibliography{bibfile}

\end{document}